\documentclass[ twocolumn ]{aastex631}
\usepackage{CJK}

\received{July, 12, 2023}
\revised{October 30, 2023}
\accepted{October 30, 2023}
\published{February 13, 2024}

\shorttitle{The Physical Properties of Changing-look Blazars}
\shortauthors{Kang et al.}

\graphicspath{{./}{figures/}}

\begin{document}
\begin{CJK*}{UTF8}{gbsn}

\title{The Physical Properties of Changing-look Blazars}


\correspondingauthor{Shi-Ju Kang, Qingwen Wu}
\email{kangshiju@alumni.hust.edu.cn; qwwu@hust.edu.cn}
\author[0000-0002-9071-5469]{Shi-Ju Kang}   
\affiliation{School of Physics and Electrical Engineering,  Liupanshui Normal University,  Liupanshui, Guizhou, 553004, People's Republic of China}
\author[0000-0001-8879-368X]{Bing Lyu} 
\affiliation{Kavli Institute for Astronomy and Astrophysics, Peking University, Beijing, Beijing, 100871, People's Republic of China}
\author[0000-0003-4773-4987]{Qingwen Wu}  
\affiliation{Department of Astronomy, School of Physics, 
Huazhong University of Science and Technology, Wuhan, Hubei, 430074, People's Republic of China}
\author[0000-0003-0170-9065]{Yong-Gang Zheng} 
\affiliation{Department of Physics, Yunnan Normal University, Kunming, Yunnan, 650092, People's Republic of China}
\author[0000-0002-5929-0968]{Junhui Fan} 
\affiliation{Center for Astrophysics, Guangzhou University, Guangzhou 510006, Peopleʼs Republic of China}
%


\begin{abstract}

Changing-look active galactic nuclei (AGNs) are a special class of AGNs that change their spectral type from type 1 to type 2 or vice versa. In recent years, a number of changing-look blazars (CLBs) were also reported, which transition between flat-spectrum radio quasars and BL Lacs. The physical properties of CLBs are still unclear. Using the $mclust$ R package for Gaussian Mixture Modeling, we performed a clustering analysis for a sample of 105 CLBs selected from the literature. Three kinds of analysis found that CLBs lie in between the parameter distributions of FSRQs and BL Lacs: (i) univariate analysis; (ii) bivariate analysis; and (iii) multivariate analysis, carried out with a dimension reduction approach of the physical properties of the three types of blazars. Our results suggest that CLBs belong to a transition type between FSRQs and BL Lacs, which may be regulated by the change of accretion process and may be similar to other changing-look AGNs.

\end{abstract}



\section{Introduction}\label{sec1}

Active galactic nuclei (AGNs) are a special class of galaxies with extremely bright nuclei, which are believed to be powered by the accretion of material into supermassive black holes. Based on the strength of emission lines, AGNs are classified as type 1, with broad lines (1000-20,000 $\rm km~s^{-1}$), or type 2, with only narrow lines (e.g., $<$ 1000 $\rm km~s^{-1}$). Blazars are a subclass of radio-loud AGNs, whose relativistic jets point to us directly \citep{1995PASP..107..803U}. Based on the strength of the optical spectral lines (e.g., the equivalent width (EW) of the spectral line is greater or less than 5 \AA), blazars are further classified as flat-spectrum radio quasars (FSRQs) with strong emission lines (EW $\ge$ 5 \AA), and BL Lacerate objects (BL Lacs) with fainter or no emission lines (EW $<$ 5 \AA) \citep{1991ApJ...374..431S,1991ApJS...76..813S}.

\vspace{2.0mm}

Changing-look AGNs (CLAGNs) are referred to as sources that experience transitions between type 1 and type 2 with significant variations of broad emission lines (see, e.g., \citealt{2023NatAs...7.1282R} for more discussions and references therein). The discovery of CLAGNs challenges the unified model of AGNs. Similar to CLAGNs, changing-look blazars (CLBs) also experience transitions between states of FSRQs and BL Lacs  (e.g., \citealt{2016AJ....151...32A};   \citealt{2021ApJ...913..146M}; \citealt{2021AJ....161..196P};  \citealt{2021Univ....7..372F,2022Univ....8..587F}, and references therein).

\vspace{2.0mm}

\citet{2021ApJ...913..146M} reported a CLB, B2 1420+32 (also named OQ 334), which experienced transitions between a FSRQ and BL Lac several times within several years. Recently, more and more CLBs have been discovered. \citet{2021AJ....161..196P} reported 26 CLBs; the authors are confident that three of these have a changing-look nature, based on optical spectra from the Large Sky Area Multi-object Fiber Spectroscopic Telescope (LAMOST) Data Release 5 (DR5) archive. In \cite{2022Univ....8..587F, 2021Univ....7..372F}, they compiled a gamma-ray jetted AGN sample based on the 4FGL catalog and reported 34 CLAGNs, 32 of which are labeled as blazars (24 FSRQs, seven BL Lacs, and one blazar candidate of unknown types) based on a significant change in optical spectral lines (see \citealt{2022Univ....8..587F,2021Univ....7..372F} for more details and references therein). \cite{2022ApJ...936..146X} reported 52 CLBs based on the EW of their optical emission lines.

\vspace{2.0mm}

The physical mechanism for the transition between an FSRQ and BL Lac is unclear. The decrease of EW for broad lines in some CLBs may be caused by the strong (beamed) jet continuum variability (e.g., \citealt{1995ApJ...452L...5V}; \citealt{2012MNRAS.420.2899G}; \citealt{2014ApJ...797...19R}; \citealt{2019RNAAS...3...92P}) or jet bulk Lorentz factor variability (e.g., \citealt{2009A&A...496..423B}) or, in some transition sources with weak radiative cooling, the broad lines are overwhelmed by the nonthermal continuum  (e.g., \citealt{2012MNRAS.425.1371G}). Furthermore,  some strong broad lines of FSRQ-type sources are missed due to a high redshift (e.g., $z > 0.7$; \citealt{2015MNRAS.449.3517D}). For instance, one of the strongest $H{\alpha}$ lines falls outside the optical window, where broad lines are not observed. Several observational effects (e.g., signal-to-noise ratio, spectral resolution, etc.)  may also affect the optical classification (see \citealt{2021AJ....161..196P} for the related discussions).

\vspace{2.0mm}

 The changing-look phenomena in blazars can provide useful insight into understanding the possible mechanism of state transition of the accretion process in supermassive black holes, the possible intrinsic variation of the jet (e.g., particle acceleration processes, jet structure etc.), and the possible accretion disk-jet connection (e.g., \citealt{2014ApJ...797...19R}; \citealt{2021AAS...23740807M}). In this work, we explored the properties for a sample of 105 CLBs and compared them with a sample of FSRQ and BL Lac objects. The sample and methods are provided in Section 2. The results are presented in Section 3. The discussion and conclusion are presented in Section 4.

\begin{table*}[hbpt!]
\caption{The Median  and  Mean of  the Parameters  for CLBs,  BL Lacs, and FSRQs}
\tiny
\centering
\begin{tabular}{ccccccccccccccccccc}
\hline \hline 
                                 &    \multicolumn{2}{c}{{FSRQ}}  && \multicolumn{2}{c}{{KS-test}}&& \multicolumn{2}{c}{{CLBs}} && \multicolumn{2}{c}{{KS-test}}&& \multicolumn{2}{c}{{BL Lacs}} && \multicolumn{3}{c}{{Source Number}}  \\ 
                                         \cline{2-3}                                   \cline{5-6}                                    \cline{8-9}    \cline{11-12}   \cline{14-15}  \cline{17-19} 
 Parameters              &      mean   &median && D   & p    && mean   &median && D   & p &&  mean   &median   &trends &{$N_{\rm B}$} & {$N_{\rm C}$} & {$N_{\rm F}$ }   \\ 
(1)                             &        (2)     &   (3)      && (4) & (5)   & &    (6)     &   (7)   && (8)  & (9)                     & & (10)  & (11)  & (12)  & (13)    & (14)  & (15)    \\ 
\hline 
$\Gamma_{\rm ph}$	   &	2.48     &	2.46 	   & & 0.48 	&  $<$1.00E-16     & &	2.26   &	2.26 	 	& &	0.55 	&   $<$1.00E-16	& &	2.02  	&	 2.01  	&	$\downarrow$   &	1397	&	105	&	748	\\
$\alpha$   	   	   	   &	2.41     &	2.40 	   & & 0.39 	&	7.59E-13       & &	2.19   &	2.19 	 	& &	0.51 	&   $<$1.00E-16	& &	1.92  	&	 1.93  	&	$\downarrow$ 	&	1397	&	105	&	748	\\
${\rm HR}_{34}$ 	   &	-0.25    &	-0.24 	   & & 0.40 	&	5.13E-13       & &	-0.09  &	-0.12 	 	& &	0.43	&	   2.22E-16 & &	0.09  	&	 0.05  	&	 $\uparrow$ 	&	1397	&	105	&	748	\\
${\rm HR}_{45}$ 	   &	-0.35    &	-0.34 	   & & 0.38 	&	3.05E-12       & &	-0.19  &	-0.20 	 	& &	0.51 	&   $<$1.00E-16	& &	0.04  	&	 0.04  	&	 $\uparrow$ 	&	1397	&	105	&	748	\\
logCD                  &	0.55     &	0.53 	   & &	0.37 	&	3.59E-10       & &	0.20   &	0.17 	 	& &	0.72 	&   $<$1.00E-16	& &	-0.48 	&	-0.48  	&	$\downarrow$ 	&	307 	&	95	&	523	\\
${\rm log}\lambda$     &	-0.92    &	-0.86 	   & &	0.57 	&  $<$1.00E-16     & &	-1.66  &	-1.59 	 	& &	0.69 	&   $<$1.00E-16	& &	-3.69 	&	-3.08 	&	$\downarrow$ 	&	307 	&	95	&	523	\\
${\rm log}L_{\rm disk}$&   45.75     &	45.84 	   & &	0.42  	&	7.10E-13 	   & &	44.88  &	44.98 	 	& &	0.60 	&   $<$1.00E-16	& &	43.22  	&	43.80 	&	$\downarrow$ 	&	307 	&	95	&	523	\\
$z$	   	   	    	   &	1.19 	 &	1.12 	   & &	0.26  	&      1.23E-05    & &	0.89   &	0.79 	 	& &	0.45 	&	   3.33E-16 & &	0.41  	&	  0.30	&	$\downarrow$ 	&	818 	&	101	&	748	\\
${\rm log}M_{\rm BH}$  &	8.55 	 &	8.57 	   & &	0.15 	&      0.05        & &	8.42   &	8.51 	 	& &	0.18 	&	    0.05    & &	8.79  	&	  8.79 	&$\searrow\nearrow$ &	307 	&	95	&	523	\\
Dir1                   &  -0.71      &   -0.74 	   & &	0.60 	& $<$1.00E-16      & &	0.02   &	-0.03   	& &	0.77 	&	$<$1.0E-16	& &	1.26  	&	  1.30 	&	$\uparrow$      &	291 	 &	94	&	523	\\
 \hline 
\end{tabular}\\
\label{Tab_median}
Note. Column 1 presents the parameters. 
Columns 2 and 3 list the mean and median of FSRQs;
columns 6 and 7 list the mean and median of CLBs;
columns 10 and 11 list the mean and median of BL Lacs, respectively.
The two-sample Kolmogorov-Smirnov test (KS-test) results between FSRQs and CLBs for the test statistic (D) and the \textit{p}-value(p) are presented in columns 4 and 5, respectively. 
The two-sample KS-test results between CLBs and BL Lacs for the test statistic (D) and the \textit{p}-value(p)  are presented in columns 8 and 9, respectively. 
Arrows are used to demonstrate trends of mean and median from FSRQs to CLBs to BL Lacs, shown in column 12.
The number of BL Lacs ($N_{\rm B}$), CLBs ($N_{\rm C}$), and FSRQs ($N_{\rm F}$) used in Figure \ref{Fig_density_01} are listed in columns 13, 14, and 15, respectively. 
Dir1 represents one-dimension reduction parameter in $mclust$ dimension reduction analysis
(e.g., see Dir1 in Figure \ref{Fig_DR_Gamma} and Section \ref{sec:Preliminary_Results}).
\end{table*}

\section{Sample and Method} \label{sec:Sample}

\subsection{Sample} \label{sec:sample}

Based on 4FGL catalogs (4LAC-DR3, \citealt{2022ApJS..263...24A}, 4LAC-DR3 \citealt{2022ApJS..260...53A}), we complied a sample of 2250 sources (1397 BL Lacs, 105 CLBs, and 748 FSRQs; see Table \ref{Tab_median}) with the gamma-ray properties gamma-ray photon index ($\Gamma_{\rm ph}$) and spectral slope ($\alpha$ at $E_0$, photon index at pivot energy when fitting with LogParabola) and the hardness ratios (e.g., see \citealt{2012ApJ...753...83A}) ${\rm HR}_{34}$ (3: 300MeV$-$1GeV; 4: 1$-$3GeV), and ${\rm HR}_{45}$ (4: 1$-$3GeV; 5: 3$-$10GeV). The hardness ratios are calculated using the following equation:
\begin{equation}\label{equ1}
HR_{ij} = \frac{\nu{F}\nu_j - \nu{F}\nu_i}{\nu{F}\nu_j + \nu{F}\nu_i}
\end{equation}
where $i$ and $j$ are indices corresponding to the different spectral energy bands defined in the 4FGL-DR3 catalog: $i, j=3$: 300MeV$-$1GeV; 4: 1$-$3GeV; and 5: 3$-$10GeV; $\nu{F}\nu$ is the spectral energy distribution over each spectral band. Among the 2250 blazars, there are 1667 (818  BL Lacs, 101 CLBs, and 748 FSRQs; see Table \ref{Tab_median}) with redshift measurements ($z$). Crossmatching the 4FGL sample (2250 blazars) and \cite{2021ApJS..253...46P} blazars sample, there are 925 common sources (307 BL Lacs, 95 CLBs, and 523 FSRQs; see Table \ref{Tab_median}). Measurements of the black hole mass ($M_{\rm BH}$), the luminosity of the accretion disk ($L_{\rm disk}$) in Eddington units ($\lambda = L_{\rm disk}/L_{\rm Edd}$), and Compton dominance (CD; the ratio of the inverse Compton to synchrotron peak luminosities) of the 925 blazars are collected from \cite{2021ApJS..253...46P}.

\vspace{2.0mm}

It should be noted that the compilation of 105 CLBs (including the CLBs or transition sources reported in previous literature) are collected from  
\cite{2022Univ....8..587F,2021Univ....7..372F};
\cite{2022ApJ...935....4Z};
\cite{2022ApJ...936..146X};
\cite{2021AJ....161..196P};
\cite{2021ApJ...913..146M};     
\cite{2016AJ....151...32A};    
\cite{2014ApJ...797...19R};
\cite{2012ApJ...748...49S};
\cite{2011MNRAS.414.2674G};
\cite{1995ApJ...452L...5V};   
\cite{2019RNAAS...3...92P};       
\cite{2014MNRAS.445.4316C};  
\cite{2019MNRAS.484L.104P};    
\cite{2009A&A...496..423B}.
All of them are compiled into an online changing-look (transition) blazars catalog (see Tables 1-9 of the TCLB Catalog,\footnote{\url{https://github.com/ksj7924/CLBCat}}, S.-J. Kang et al. 2024a, in preparation), which is archived on Zenodo:10.5281/zenodo.10061349,\footnote{\url{https://www.zenodo.org/record/10061349}} (\citealt{2023kang}) for easy communication.

\subsection{Mclust and \textit{UMAP}} \label{sec:mclust}

The $mclust$ R package is employed to calculate the univariate density distribution of some parameters and the multivariate dimension reduction of some parameters. \textit{Mclust} is a popular R package complied by \cite{mclust_2016,Scrucca_book} using R Language \citep{R_code} for model-based clustering, classification, and density estimation based on finite Gaussian mixture modeling. This package also provides several tools for model-based clustering, classification, and density estimation, which include Bayesian regularization and dimension reduction using the expectation-maximization (EM) algorithm (e.g.,  \citealt{mclust_2016,Scrucca_book}).

\vspace{2.0mm}

Uniform manifold approximation and projection (UMAP) is a dimension reduction technique that is used for visualization similarly to t-stochastic neighbourhood embedding (t-SNE; \citealt{tsne2008}), which visualizes high-dimen-sional data by giving each data point a location in a two- or three-dimensional map, and is also employed for dimension reduction (\citealt{2018arXiv180203426M}; \citealt{umap_R}).

\section{Results}\label{sec:Preliminary_Results}

\begin{figure*}[htp!]
\centering
\includegraphics[width=16cm,height=12cm]{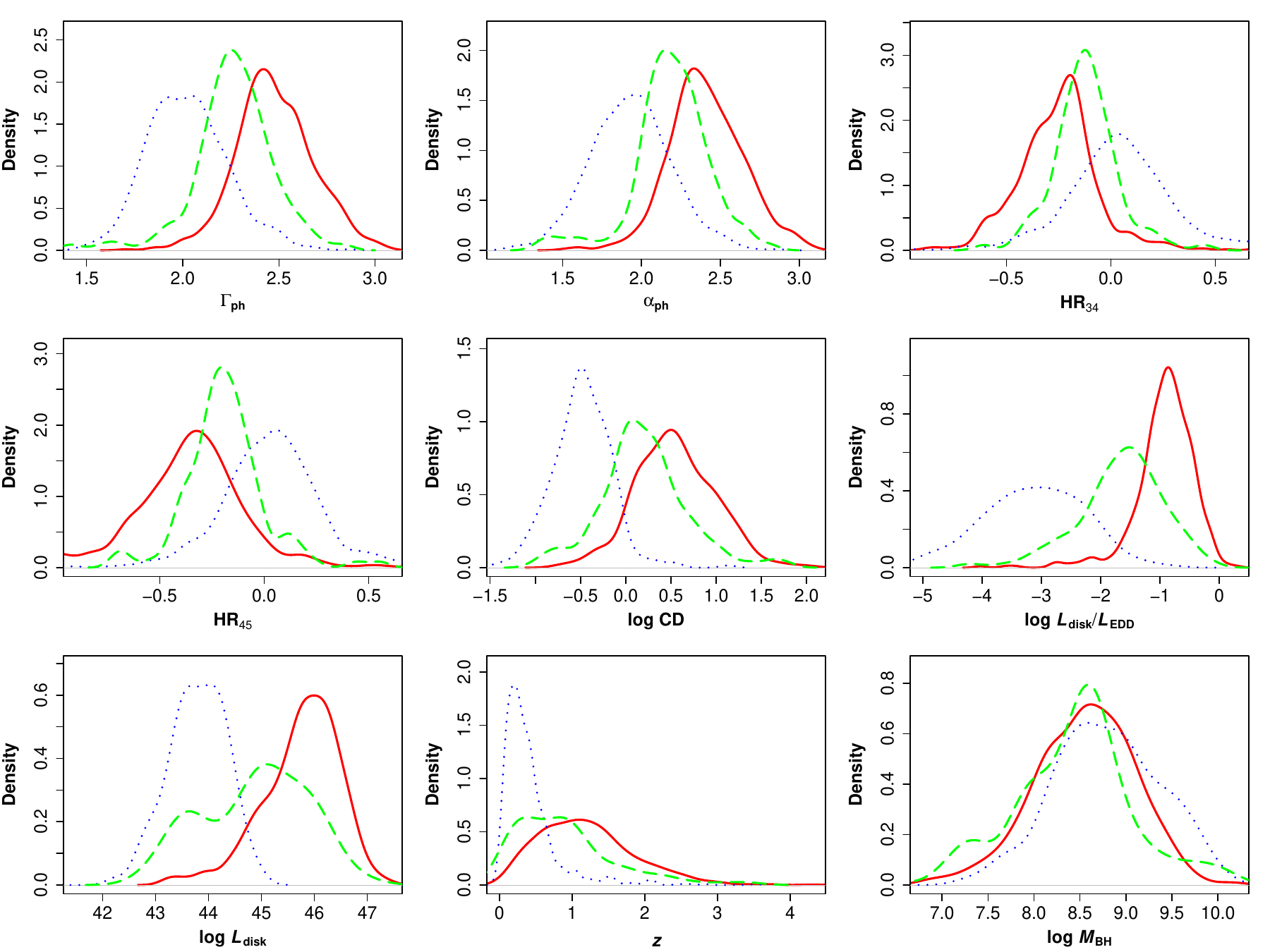}
\caption{The density distributions of 
the gamma-ray photon spectral index ($\Gamma_{\rm ph}$),
the spectral slope ($\alpha_{\rm ph}$ at $E_0$, photon index at pivot energy when fitting with LogParabola), 
the hardness ratios ${\rm HR}_{34}$ (3:300 MeV$-$1 GeV; 4:1$-$3 GeV),
the hardness ratios ${\rm HR}_{45}$ (4:1$-$3 GeV; 5: 3 $-$ 10 GeV),
the CD (the ratio of the inverse Compton to synchrotron peak luminosities),
the accretion disk luminosity ($L_{\rm disk}$) in Eddington units ($L_{\rm disk}/L_{\rm Edd}$),
the accretion disk luminosity ($L_{\rm disk}$),
the redshift ($z$),
and
the black hole mass ($M_{\rm BH}$).
The blue dotted lines, red solid lines, and green dashed lines  
represent the BL Lacs, FSRQs, and CLBs respectively. 
\label{Fig_density_01}}
\end{figure*}

Using the $densityMclust()$ function of the $mclust$ R package for density estimation, the density distributions of $\Gamma_{\rm ph}$, $\alpha_{\rm ph}$, ${\rm HR}_{34}$, and ${\rm HR}_{45}$ for 2250  blazars (1397 BL Lac, 105 CLBs, and 748 FSRQs; see Table \ref{Tab_median}), the $CD$, $\lambda$=$L_{\rm disk}/L_{\rm Edd}$, $L_{\rm disk}$, $M_{\rm BH}$ for 925 blazars (307 BL Lac, 95 CLBs, and 523 FSRQs), and the $z$ for 1667 blazars (818 BL Lac, 101 CLBs, and 748 FSRQs) are shown in Figure \ref{Fig_density_01}, where the blue dotted, red solid, and green dashed lines represent the density distributions for BL Lacs, FSRQs, and CLBs, respectively. Comparing the density distributions of CLBs with those of BL Lacs and FSRQs in Figure \ref{Fig_density_01}, the density distributions of CLBs are obviously located between those of BL Lacs and those of FSRQs for $\Gamma_{\rm ph}$, $\alpha_{\rm ph}$,  ${\rm HR}_{34}$, ${\rm HR}_{45}$, CD, $L_{\rm disk}$, and $\lambda$=$L_{\rm disk}/L_{\rm Edd}$. The median and mean values (see Table \ref{Tab_median}) of $\Gamma_{\rm ph}$, $\alpha$, logCD,  log$L_{\rm disk}$, and ${\rm log}\lambda$ of CLBs are larger than those of BL Lacs, and less than those of FSRQs, respectively. Meanwhile, the median and mean for the  ${\rm HR}_{34}$, and ${\rm HR}_{45}$ of CLBs are less than those of BL Lacs and are greater than those of FSRQs, respectively. The two-sample Kolmogorov-Smirnov test (KS-test) for these parameters gives the value of the test statistic $D$ $\geq$ 0.37 with a \textit{p}-value of $p \leq 3.59\times10^{-10}$, where $p$ $>$ 0.05 indicates that the two populations should be the same distribution. This result indicates that the distributions of these parameters are independent between CLBs and FSRQs or between CLBs and BL Lacs, respectively.

\begin{figure*}[htp!]
\centering
\figurenum{2}
\includegraphics[width=18cm,height=15cm]{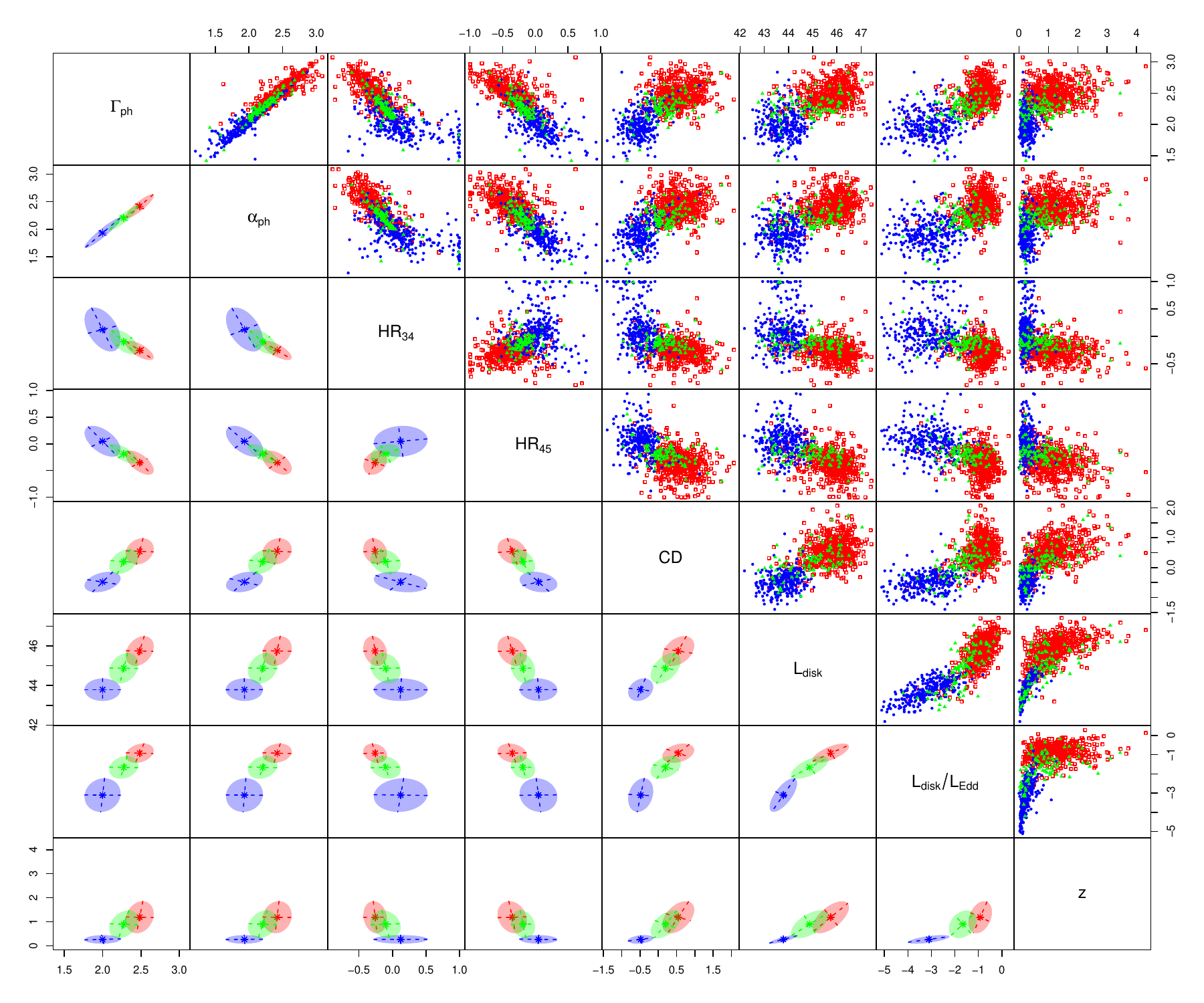}
\caption{Scatterplots of 
the gamma-ray photon spectral index ($\Gamma_{\rm ph}$),
the spectral slope ($\alpha_{\rm ph}$ at $E_0$, photon index at pivot energy when fitting with LogParabola), 
the hardness ratios ${\rm HR}_{34}$ (3:300 MeV$-$1 GeV; 4:1$-$3 GeV),
and ${\rm HR}_{45}$ (4:1$-$3 GeV; 5:3$-$10 GeV),
redshift ($z$),
the CD (the ratio of the inverse Compton to synchrotron peak luminosities),
the accretion disk luminosity ($L_{\rm disk}$),
and
the accretion disk luminosity ($L_{\rm disk}$) in Eddington units ($L_{\rm disk}/L_{\rm Edd}$),
which are shown in the upper-right panels,
where the red hollow squares, blue solid dots, and green triangles
indicate the FSRQs, BL Lacs,  and CLBs respectively.
The bottom-left panels (mirroring the upper-right panels) represent the best-fitting values and 1 $\sigma$ confidence regions that are obtained from a Gaussian fitting each pairwise parameter for the three subsamples separately, which are marked as crosses and ellipse regions, respectively.
\label{Fig_Scatterplots_Gamma}}
\end{figure*}

\vspace{2.0mm}

It should be noted that the density distributions of CLBs for redshift showed a wide distribution, where the mean (0.89) and median (0.79) of redshift for CLBs are greater than those of BL Lacs (mean of 0.41 and median of 0.30) and are less than those of FSRQs (mean of 1.19 and median of 1.12). However, the distributions are not much different for redshift for CLBs and FSRQs, where the density distribution of CLBs for redshift is similar to that of FSRQs (see Figure \ref{Fig_density_01}). The KS-test gives $D = 0.26$ with a \textit{p}-value of $p =  1.23 \times 10^{-5}$, while the KS-test between CLBs and BL Lacs gives $D = 0.45$ with a \textit{p}-value of $p =  3.33 \times 10^{-16}$, which indicates that the distributions of redshift are independent between CLBs and BL Lacs. However, it is a fact that it is very hard to measure the redshift of BL Lacs at a high redshift, which is a possible bias of the distribution in Figure \ref{Fig_density_01}. In addition, we found that the distributions of black hole mass (${\rm log}M_{\rm BH}$) are similar for FSRQs, BL Lacs, and CLBs, where  $p \simeq 0.05$ for CLBs/FSRQs and CLBs/BL Lacs, respectively, based on the KS-test.

\vspace{2.0mm}

For the parameters $\Gamma_{\rm ph}$, $\alpha_{\rm ph}$, ${\rm HR}_{34}$, ${\rm HR}_{45}$, CD, $L_{\rm disk}$, $\lambda$=$L_{\rm disk}/L_{\rm Edd}$, and $z$, the univariate density distributions of CLBs are located between those of BL Lacs and those of FSRQs. The scatterplots of these parameters (for each pair of parameters with valid values of data; for instance, there are 2250 valid data points in the $\Gamma_{\rm ph}$-${\rm HR}_{45}$ plot, 925 valid data points in the $\Gamma_{\rm ph}$-CD plot, respectively, etc.) are presented in the upper-right panels of Figure \ref{Fig_Scatterplots_Gamma}, where the red hollow squares, blue solid dots, and green triangles represent FSRQs, BL Lacs, and CLBs, respectively. Based on a Gaussian fitting to each pairwise parameter for the three subsamples separately, the best-fitting values and 1$\sigma$ confidence regions are obtained and marked as crosses and ellipse regions in the bottom-left panels (which mirror the upper-right panels). We note that the green triangle CLBs are mainly scattered at the middle boundary between FSRQs and BL Lacs.

\vspace{2.0mm}

\begin{figure*}[htp!]
\centering
\figurenum{3}
\includegraphics[width=16cm,height=8cm]{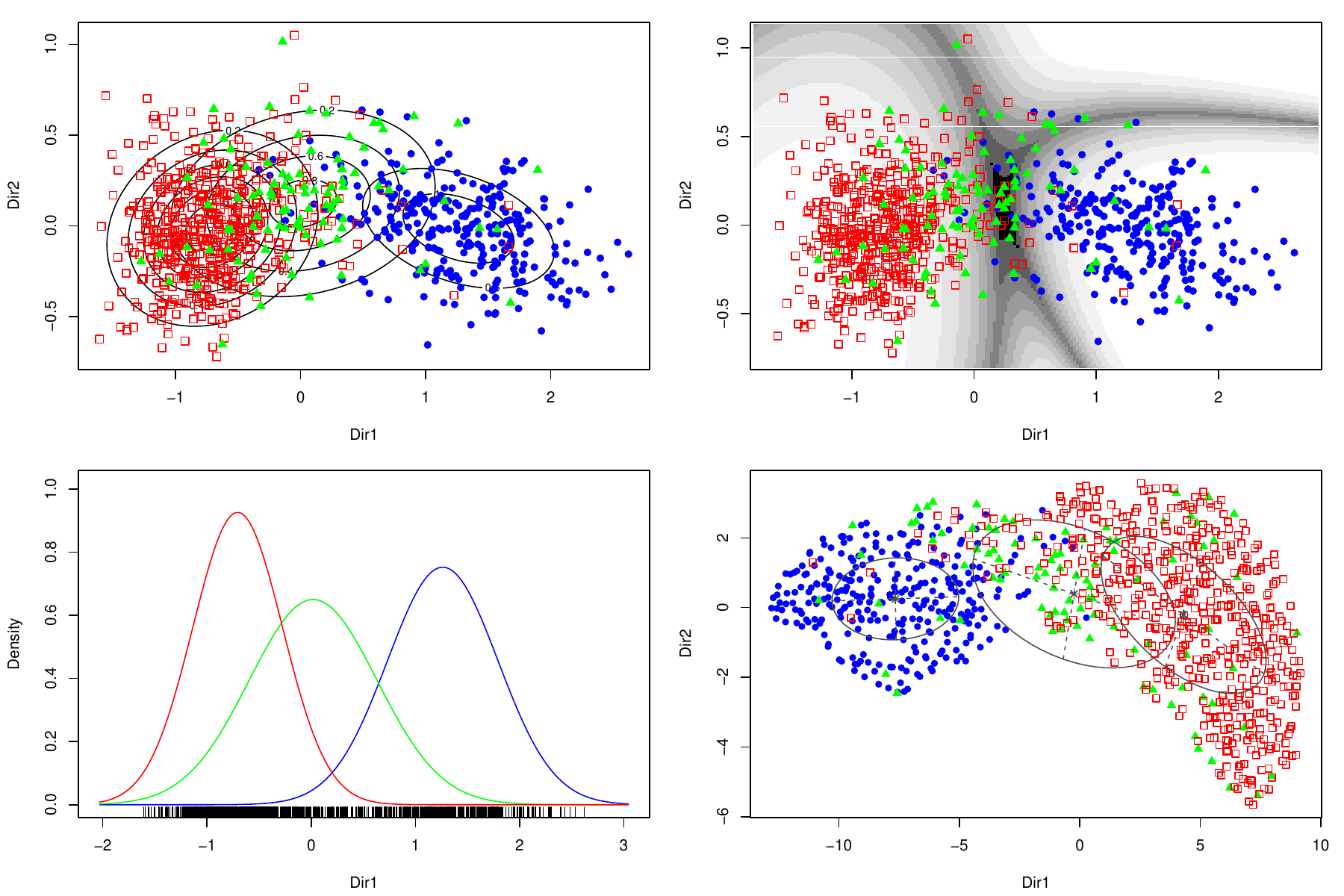}	
\caption{Dimension reduction for model-based clustering and classification (\textit{mclustDR}, first row and bottom-left panels),
and UMAP (bottom-right panel),
the multidimensional data ($\Gamma_{\rm ph}$,  $\alpha_{\rm ph}$, 
 ${\rm HR}_{34}$, ${\rm HR}_{45}$,
CD, $L_{\rm disk}$,  $\lambda$ = $L_{\rm disk}/ L_{\rm Edd}$, and $z$) 
with three groups (BL Lacs, FSRQs and CLBs) are dimensionally reduced to two dimensions,
where the red hollow squares, blue solid dots, and green triangles
indicate the BL Lacs, FSRQs, and CLBs, respectively.
The grey color map in the top-right panel shows the uncertainty boundaries of the mixture densities.}
\label{Fig_DR_Gamma}
\end{figure*}

The \textit{mclustDR}() function\footnote{\url{https://rdrr.io/cran/mclust/man/MclustDR.html}} in the $mclust$ R package  (\citealt{mclust_2016}) is a dimension reduction method for visualizing the clustering or classification structure obtained from a finite mixture of Gaussian densities, which aims at reducing the dimensionality by identifying a set of linear combinations, ordered by importance as quantified by the associated eigenvalues, of the original features, which capture most of the clustering or classification structure contained in the data. Using this R function \textit{mclustDR}(), a dimension reduction is performed to the multidimensional data ($\Gamma_{\rm ph}$,  $\alpha_{\rm ph}$,  ${\rm HR}_{34}$, ${\rm HR}_{45}$, CD, $L_{\rm disk}$, $\lambda$ = $L_{\rm disk}/ L_{\rm Edd}$, and $z$) with three groups, where there are only 908 blazars with 291 BL Lacs, 94 CLBs, and 523 FSRQs for all eight parameters with valid data. The dimension reduction performs a projection on a two-dimensional subspace, making use of linear combinations of the eight original features of each subsample. In the case of three groups in two dimensions, the CLB sources  (green)  are also located between FSRQs  (red)  and BL Lacs (blue) (see Figure \ref{Fig_DR_Gamma}). The contour and density distributions of the dimension reduction parameters for CLBs are obviously located between those of FSRQs and BL Lacs, where the mean values are 0.02, -0.71, and 1.26, respectively, for the one-dimensional reduction parameter (e.g., Dir1 in Figure \ref{Fig_DR_Gamma}). The two-sample KS-test for Dir1 is $D$ = 0.60 with a \textit{p}-value of $p < 1.00\times10^{-16}$ between CLBs and FSRQs. For CLBs and BL Lacs, the test gives $D$ = 0.77 with a \textit{p}-value of $p <  1.00 \times 10^{-16}$ (see Table \ref{Tab_median}). In order to further test the results of the $mclust$ dimension reduction analysis, the UMAP algorithm \citep{2018arXiv180203426M,umap_R} is applied to the multidimensional data of the same eight parameters. The same dimension reduction results are presented, where the CLBs (green) are also located between FSRQs (red) and BL Lacs (blue) (see the bottom-right panel in  Figure \ref{Fig_DR_Gamma}).

\vspace{2.0mm}

The results of univariate, bivariate, and multivariate analysis indicate that the CLBs are in an intermediate transition state between FSRQs and BL Lacs.  This suggests that CLBs may be a class of excessive blazar sources in transition between FSRQs and BL Lacs.

\section{Discussion and Conclusion} \label{sec:discussion_conclusion}

In this article, we first systematically explored the statistical properties of CLBs based on the 4FGL catalogs (4FGL-DR2, \citealt{2022ApJS..260...53A}; 4LAC-DR3, \citealt{2022ApJS..263...24A}) and \cite{2021ApJS..253...46P}'s blazars sample. We found that there are eight parameters: $\Gamma_{\rm ph}$, $\alpha_{\rm ph}$,  ${\rm HR}_{34}$, ${\rm HR}_{45}$, CD, $L_{\rm disk}$, $\lambda$ = $L_{\rm disk}/ L_{\rm Edd}$, and $z$. The univariate density distributions of CLBs are located between those of BL Lacs and those of FSRQs. The scatterplots of these eight parameters display a similar distribution, whereby the CLBs are mainly scattered in the middle region between FSRQs and BL Lacs. Using the dimension reduction function $mclustDR()$ on the eight parameters, the multidimensional data of three groups are dimensionally reduced to two dimensions. In the case of the three groups in two dimensions, the CLB sources are also located at the cross districts between FSRQs and BL Lacs. These results suggest that CLBs most possibly stay in a transition state between FSRQs and BL Lacs.

\vspace{2.0mm} 

Currently, it is generally believed that the dichotomy between FSRQs and BL Lacs is caused by different accretion modes (see, e.g., \citealt{2019MNRAS.482L..80B,2019MNRAS.486.3465M,2021MNRAS.505.4726K,2022Galax..10...35P}  for more details and references therein), where the standard thin disk exists in FSRQs while radiatively inefficient advection-dominated accretion flow (ADAF) exists in BL Lacs (e.g.,  \citealt{2014ARA&A..52..529Y}). The transition between the standard cold accretion disk and the ADAF disk may occur at a critical Eddington ratio around several percent (see, e.g., \citealt{1998tbha.conf..148N,2016ApJ...817...71C} for more details and references therein). It should be noted that the median and mean Eddington-scaled accretion disk luminosity of CLBs are $\lambda$=$L_{\rm disk}/L_{\rm Edd}$$\simeq$ -1.59 and -1.66, respectively, which roughly correspond to the critical Eddington ratios for the transition of the accretion mode. This is also consistent with the Eddington ratios in log($L_{\rm disk}/L_{\rm Edd}$) = -2.70 $\sim$ -1.07 for a potential transition between BL Lacs and FSRQs, suggested by \cite{2022ApJ...925...97P}. This further indicates that CLBs are in an intermediate transition state between FSRQs and BL Lacs, and CLBs are a class of excessive blazar source in transition from FSRQs to BL Lacs.

\vspace{2.0mm} 

\vspace{2.0mm} 

We find that there are eight parameters: $\Gamma_{\rm ph}$, $\alpha_{\rm ph}$, ${\rm HR}_{34}$,  ${\rm HR}_{45}$, CD, $L_{\rm disk}$, $\lambda$=$L_{\rm disk}/L_{\rm Edd}$, and $z$. The density distributions of CLBs are located between those of BL Lacs and those of FSRQs. These properties can be used to search for more CLB candidates (S. -J. Kang et al. 2024b, in preparation). The five parameters $\Gamma_{\rm ph}$, $\alpha_{\rm ph}$, ${\rm HR}_{34}$,  ${\rm HR}_{45}$, and CD are all gamma-ray parameters or properties of associated gamma rays, which may be not essential and may be modulated by the jet or/and accretion disk. For instance, FSRQs and BL Lacs may have different accretion modes (see, e.g.,  \citealt{2019MNRAS.482L..80B,2019MNRAS.486.3465M,2021MNRAS.505.4726K,2022Galax..10...35P} for more details and references therein), where FSRQs have a standard cold accretion disk and BL Lacs have an ADAF (e.g.,  \citealt{2014ARA&A..52..529Y}). Different accretion modes provide different environment for jet Comptonization (e.g., different external seed photons) (e.g., \citealt{1998MNRAS.301..451G,2016Galax...4...36G,2017MNRAS.469..255G,2022Galax..10...35P}), where the cooling of relativistic electrons should be different (\citealt{1998MNRAS.301..451G}). FSRQs normally have a standard cold accretion disk and strong broad emission lines and torus, which provide a fast-cooling environment for the external Compton process, and high-energy gamma-ray radiation is dominated by the external Compton process. Therefore, the inverse Compton process becomes dominant with respect to the synchrotron emission, which implies a larger Compton dominance. However, disk emission, broad emission lines, and torus are weak or absent in BL Lacs; the lack of ambient seed photons implies that the synchrotron self-Compton process dominates the high-energy gamma-ray emission (see, e.g., \citealt{1998MNRAS.301..451G,2022Galax..10...35P} for more discussions and references therein), which leads to a small Compton dominance. Therefore, these five gamma-ray (or associated) parameters may not be essential. The difference between CLBs, FSRQs, and BL Lacs may be mainly regulated by the evolution of the accretion process, which is similar to that of other CLAGNs (e.g., \citealt{2022ApJ...927..227L}; \citealt{2023NatAs...7.1282R} and references therein).  For instance,  the distribution of accretion disk luminosity in Eddington units ($\lambda = L_{\rm disk}/L_{\rm Edd}$) for CLBs is located between those of FSRQs and BL Lacs. This is similar to the Eddington-scaled mid-IR luminosities of CLAGNs that stay between those of low-luminosity AGNs and high-luminosity AGNs \citep{2022ApJ...927..227L}, which may be regulated by the change of accretion process and similar to other CLAGNs. Whether they have similar transition stage roles with other CLAGNs needs further investigation.

\vspace{2.0mm} 

In the CLB OQ 334, there is also evidence for the transition of accretion modes. CLB OQ 334 stayed in a stable state for a long time before the EW change and state transition. After state transitions and EW changes, the variability of gamma-ray bands also begins to be accompanied by drastic changes, which indicates that they are related. Using a one-zone lepton jet model, the multiband spectral energy distribution of OQ 334 is simulated. It was found that the gamma-ray variability may be caused by changes in the abundance of external soft photons due to changes in the luminosity of the accretion disk (\citealt{ren2024fermilat}). We argue that the change in luminosity of the accretion disk may be essentially caused by the change in accretion of the central object (black hole). The accretion of the central object black hole may play an important role in the transition of the state of CLB OQ 334.

\vspace{1.5mm} 

In the case of the three groups of CLBs, FSRQs, and BL Lacs, the CLBs are in an intermediate transition state between FSRQs and BL Lacs and may be a class of excessive blazar sources in transition from FSRQs to BL Lacs. The accretion of the central object black hole may play an important role in the state transition of the CLBs between FSRQs and BL Lac. However, it should be noted that CLBs are small viewing angle jetted AGNs. The jet effect (e.g., \citealt{1995ApJ...452L...5V}; \citealt{2009A&A...496..423B}; \citealt{2012MNRAS.425.1371G}; \citealt{2012MNRAS.420.2899G}; \citealt{2014ApJ...797...19R}; \citealt{2019RNAAS...3...92P}; \citealt{2021ApJ...913..146M}) remains an unavoidable mainstream explanation (albeit still the dominant explanation) for jetted CLBs, which may be misclassified based on EW classification; for instance, some FSRQs are misclassified as BL Lacs (see \citealt{2023MNRAS.525.3201K} for related discussions and references therein). The evolution of the accretion and jet may also play a key role in CLBs, which needs to be further studied.

\vspace{2.0mm} 

In addition, we note that the density distributions of $L_{\rm disk}$ for CLBs seems to show a bimodal distribution (see Figure \ref{Fig_density_01}). Using the R $mclust$ package (\citealt{mclust_2016,Scrucca_book}) to perform Gaussian finite mixture model clustering analysis, there are two components that are the optimal number of mixture components obtained from the corresponding optimal mixture model estimation (according to Bayesian information criteria (BIC)), where a corresponding optimal BIC of -273.12 is obtained that is fitted by an $mclust$ V (univariate, unequal variance) model. In order to test and confirm the bimodal distribution, the Kaye's Mixture Model (KMM) test (see \citealt{1994AJ....108.2348A} for details and references therein) is employed to check the distribution of $L_{\rm disk}$. The KMM test showing a KMM-test probability of $p = 0.49 \times 10^{-4}$ also indicates a bimodal distribution in the $L_{disk}$ for the CLBs sample. The bimodal distribution of the $L_{\rm disk}$ for the CLBs seems to suggest that the accretion process plays an important role in the state transitions between FSRQs and BL Lacs rather than the jet effects. However, a similar dichotomy is not presented in the density distribution of $\lambda$=$L_{\rm disk}/L_{\rm Edd}$. The luminosity of the accretion disk is a bimodal distribution, but the accretion rate does not change much and does not show a bimodal distribution. This implies that it may be evidence of the transition in CLBs caused by a change of the radiation efficiency of the accretion disk (see, e.g., \citealt{2014ARA&A..52..529Y}; \citealt{2023NatAs...7.1282R} for related discussions and references therein). Considering the distributions of ${\rm log}M_{\rm BH}$ are undifferentiated, which perhaps is not to account for efficiency bimodal distribution, and assuming that small black holes correspond to small luminosity, the final Eddington ratio may not have a bimodal distribution. Whether the radiation efficiency is related to the transformation of state still needs further study.

\vspace{2.0mm} 
 
We should note that the $mclust$ clustering algorithm (R package; \citealt{mclust_2016}) and UMAP algorithm (\citealt{umap_R}) do not account for errors in the data. In addition, not all variables have uncertainty, and their primary results are not affected by variable uncertainty. For the $L_{\rm disk}$ (and $\lambda$ = $L_{\rm disk}/ L_{\rm Edd}$), some sources only have upper limits, which also does not affect the main outcome. So the uncertainties or upper limits of the measured quantities are not considered and used in the analysis. It should also be noted that CLB sources are still limited, and the sample selection effects should be considered, which may affect the source distributions and the results of the analysis. Future large-scale spectroscopic or photometric surveys will discover more CLBs. In the future, in-depth research will be of great significance to reveal the nature of CLB state transformation.
This issue will continue to be addressed in future work.

\vspace{2.0mm} 

CLBs are not only essential for studying and understanding the different properties of FSRQs and BL Lacs (e.g., ``blazar sequence"; \citealt{1998MNRAS.299..433F,2015MNRAS.450.3568X,2017ApJS..229...21X}), but also they have important implications for their redshift evolution. We note that the density distributions of CLBs for redshift showed a wide distribution (see Figure \ref{Fig_density_01}). Although not obvious, overall the distribution of CLBs is still halfway between FSRQs and BL Lacs. The median and mean of redshift for CLBs are greater than those of BL Lacs and are less than those of FSRQs (see Table \ref{Tab_median}), where the KS-test gives the value of the test statistic $D$ $\geq$ 0.26 with a $p$-value of $p \leq 1.23\times10^{-5}$ for redshift between CLBs and FSRQs, or between CLBs and BL Lacs. This may be implying that there is a redshift evolution sequence from higher redshift FSRQ to medium redshift CLBs to lower redshift BL Lacs, where the redshift value of CLBs is in the transition phase between those of FSRQs and BL Lacs. These results support the negative redshift evolution in blazars  (e.g., \citealt{2014ApJ...780...73A} and references therein). FSRQs are relatively young active galaxies with a higher redshift (FSRQs are at a higher redshift compared to BL Lacs), while BL Lacs with a lower redshift may represent the elderly and evolved phase of a blazar's life. FSRQs evolve to BL Lacs, suggesting that a BL Lac is the late stage of an FSRQ. CLBs are in an intermediate transition state between FSRQs and BL Lacs and may be a class of excessive sources in transition from FSRQs to BL Lacs. However, recently, the discovery of the high-redshift BL Lacs 4FGL J1219.0$+$3653, at a redshift of $z \sim 3.59$ (\citealt{2020ApJ...903L...8P}) and the high-redshift BL Lacs FIRSTJ233153.20+112952.1, at a redshift of $z \sim 6.57$  (\citealt{2022ApJ...929L...7K}), has greatly challenged the evolutionary scenario of blazars, whereby BL Lacs are believed to represent the final evolutionary phase of FSRQs (e.g., \citealt{2014ApJ...780...73A} and references therein).

\vspace{2.0mm} 

Furthermore, it should also be noted that the observed effect of redshift should also be considered. For instance, it is very hard to measure the redshift of BL Lacs at a high redshift, which is a possible bias of the distribution for BL Lacs. The lack of more high-redshift BL Lacs, where only 11 BL Lacs have redshift values greater than 2 and one greater than 3 in our sample, may significantly affect the distribution of BL Lacs and the evolutionary scenarios of the blazar redshift evolution sequence (e.g., \citealt{2002ApJ...564...86B,2014ApJ...797...19R,2022Galax..10...35P} and references therein). Future large spectroscopic or photometric surveys will provide a large redshift sample of blazars (and CLBs) to further test the blazar redshift evolution sequence and the role of CLBs.

\section*{Acknowledgements}

We thank the editor and anonymous referee for very constructive and helpful comments and suggestions, which greatly helped us to improve our paper. This work is partially supported by the National Natural Science Foundation of China (grant Nos. 12163002, U1931203, U2031201, and 12363002, 12233007), the National SKA Program of China (grant No. 2022SKA0120101), and the Discipline Team of Liupanshui Normal University (grant No. LPSSY2023XKTD11).


%

\facility{Fermi (LAT)}


\software{R \citep{R_code},  $mclust$ \citep{mclust_2016}, UMAP \citep{umap_R}.   }






\bibliography{aastex631}{}
\bibliographystyle{aasjournal}



\end{CJK*}
\end{document}